\documentclass[onecolumn,showpacs,floats,floatfix,aps,prl,groupedaddress]{revtex4}
\usepackage{graphicx}
\usepackage{url}
\usepackage{amsmath}
\usepackage{subfigure}

\begin{document}

\title{Universality and Properties of Neutron Star Type I Critical Collapses}

\author{Mew-Bing Wan}

\affiliation{Asia-Pacific Center for Theoretical Physics, Pohang, Gyeongbuk 790-784, South Korea} 

\date{\today}

\begin{abstract}
We study the neutron star axisymmetric critical solution previously found in numerical studies of neutron star mergers.
Using neutron star-\textit{like} initial data and performing similar merger simulations, 
we demonstrate that the solution is indeed a semi-attractor on the 
threshold plane separating the basin of a neutron star and the basin of 
a black hole in the solution space of the Einstein equations.
In order to explore the extent of the attraction basin of the neutron star semi-attractor, 
we construct initial data phase spaces for these neutron star-\textit{like} initial data.
From these phase spaces, we also observe several interesting dynamical scenarios where 
the merged object is supported from prompt collapse.
The properties of the critical index of the solution, in particular, 
its dependence on conserved quantities, are then studied. From the study, it is found that a family of  
neutron star semi-attractors exist that can be classified by both their rest masses and ADM masses. 


\end{abstract}
\maketitle

\section{I. Introduction}

Critical phenomena in gravitational collapse were discovered by Choptuik 
almost a decade ago \cite{Choptuik}. Since then, there has been much 
development in establishing the theory in its mathematical aspects. Past 
studies have considered the assumptions of both spherically 
symmetric and axisymmetric systems. Several matter models, e.g., perfect fluids, real scalar fields, 2D Sigma models, SU(2) 
Yang-Mills and Skyrme models, as outlined in a comprehensive review by Gundlach and Mart\'{i}n-Garc\'{i}a \cite{Gundlach}, 
as well as neutron stars \cite{Novak,Noble07,Noble} and black holes \cite{Pretorius,Sperhake}, were considered. 
The neutron star (NS) models studied in \cite{Noble07,Noble} employ spherical symmetry. 

The first axisymmetric simulations of NS critical gravitational collapses 
involving two colliding non-rotating NSs were performed by Jin and Suen \cite{Jin}.
The axisymmetry in these studies enables high-resolution finite differencing that facilitates the 
observation of the fine structure of critical phenomena. In Ref. \cite{Jin}, the 
central densities, the boost velocities, the separations, and the polytropic index of the equation of state (EOS) of the colliding NSs initial data,  
are varied respectively. Varying solely the central densities of the NSs corresponds to varying the 
baryonic masses of the NSs. Configurations with supercritical baryonic masses result in black holes with finite masses while those with subcritical 
baryonic masses result in NSs. The system is carried away from the critical 
solution by its one unstable mode. The time of departure from the critical solution scales with respect to constant 
powers of the distance of the initial data from the critical threshold. 
With the presence of an overall length scale in this system, and consequently the formation of black holes with finite masses, 
the critical phenomena exhibited are classed as Type I.  

However, as noted in the review by Gundlach and Mart\'{i}n-Garc\'{i}a \cite{Gundlach}, the families of NS initial data constructed by varying the boost velocities, initial separations and the adiabatic indices of the polytropic EOS, lie very close to each other and are within the perturbative region of the NS critical solution. 
As such, the universality of the NS critical solution with respect to different families of initial data is still considered unclear. 
In the current paper, we show this universality using more \textit{general} NS-\textit{like} systems as initial data.
Specifically, we construct axisymmetric packets of matter with Gaussian baryonic density distributions
with various parameters that can be varied respectively. At the initial time, these packets are characterized using the polytropic EOS,
$p=\kappa\rho^{\Gamma}=\rho\epsilon(\Gamma-1)$ with $\rho$ being the baryonic density, $\epsilon$ the specific internal energy density of the fluid, $\Gamma=2$ and $\kappa=0.0298c^2/\rho_n$, where $\rho_n=2.3\times 10^{14}g/cm^3$ is the nuclear density. 
Such an EOS would produce a neutron star that is less stiff than the Lattimer-Swesty EOS with a lepton to baryon ratio of $0.1$ \cite{Evans}.   
Since all parameters and variables in our simulations are set in geometric units, i.e., $G=c=M_{\odot}=1$, 
with the aforementioned EOS, $\kappa$ has the dimensions of $\rho^{-1}$ and the abovementioned value corresponds to $\kappa=80$ in the simulation.
With $\kappa\neq 1$, the system considered here possesses a fundamental length scale, similar to that considered in Ref. \cite{Jin}. 
The maximum baryonic mass allowable by this EOS for a non-rotating equilibrium NS is $1.61$ in the abovementioned units.
During the evolution, similar to Ref. \cite{Jin}, only the ideal gas EOS $p=\rho\epsilon(\Gamma-1)$ is used,
thus allowing for shock heating.  
As in Ref. \cite{Jin}, the packets here too collide head-on with each other under an imposed boost velocity.

From our simulations, the NS critical solution found in Ref. \cite{Jin} is observed to attract the evolutions of these different initial data.
Seen from a dynamical systems point of view, the NS critical solution lives on the threshold plane
separating the basin of a NS and the basin of a black hole in the solution space of the Einstein equations.
It attracts the evolutions of these various initial data via infinite decaying modes along the direction of the
threshold plane and repels them from the threshold plane via its one unstable mode. 
Therefore, in the dynamical systems picture, the NS critical solution is a semi-attractor.
A natural question that follows is then: If it is a semi-attractor, what is the extent of the attraction basin of the NS critical solution?
Also, what are the properties of its critical index, in particular, its dependence on the conserved quantities of the system? 
  
We outline the paper as follows. Section 2 details the methods used in our numerical simulations, and presents the universality of the
NS critical solution found in Ref. \cite{Jin} with respect to the above-mentioned families of initial data.
In Sec. 3, initial data phase spaces of both the NS systems studied in Ref. \cite{Jin} and the new NS-like systems introduced in our current paper, 
are constructed to provide further information on the properties of the NS critical solution in the solution space of the Einstein field equations. 
Using these initial data phase spaces, we determine the extent of the attraction basin of the NS critical solution in terms of several 
initial data parameters. In Sec. 4, we study the properties of the critical index of the NS critical solution.
A main feature of this section is the investigation of the dependence of the critical indices of both NS systems and NS-like systems on the conserved quantities of the systems, i.e., their rest masses and ADM masses. 
This property of the critical index will be able to shed light on the classes of NS semi-attractors that could exist.
Section 5 sums up the findings of the paper. 

\section{II. Universality}

We present in this section new initial data where the matter field consists of packets of matter whose baryonic densities are 
characterized by axisymmetric Gaussian distributions and a polytropic EOS described in the previous section. 
The baryonic density distribution of each packet of matter follows the following relation:
\begin{equation}
\rho(r_{iso})=Pe^{r^2_{iso}/2Q},
\end{equation}
where $r_{iso}=(x^2+y^2+z^2)^{1/2}$,
and $P$ and $Q$ can be freely adjusted to produce various 1-parameter families of initial data
that lie outside the perturbative region of the NS critical solution. 
For the test $3$-metric of each of the matter packets, we use a similar Gaussian function that is asymptotically flat,
and whose heights and widths can be similarly adjusted to yield different initial spacetimes. 
These height and width parameters of the spacetime and matter configuration constitute additional degrees of freedom 
for the Gaussian packet system as compared to the NS system.
This is due to the fact that the NS configuration is uniquely defined by the central densities of the two NSs
via the TOV (Tolman-Oppenheimer-Volkoff) equations, whereas the Gaussian packet configuration is not. 

The numerical setup used here is similar to that in Ref.~\cite{Jin}. 
On a thin slab with a width of $5$ grid points along the $xz$-plane of a Cartesian grid, we construct two of the aforementioned matter packets with the same mass separated at a distance from each other along the $z$ direction.
To do this, symmetry conditions are applied using the Cartoon method \cite{Alcub}.  
The matter packets are boosted toward each other in a head-on collision along the $z$ direction.
Employing York's formulation \cite{York}, we solve the Hamiltonian and momentum constraints to obtain the initial data. 
Specifically, the $3$-metric and extrinsic curvature of the two boosted Gaussian packets of matter are superimposed in the same manner as in Ref.~\cite{MillerSuenTobias}. Setting the extrinsic curvature scalar to zero, the resulting matter and momentum distributions, the conformal part of the metric, and the transverse traceless part of the extrinsic curvature are then input into York's procedure to solve for the initial data.  
The initial data obtained satisfy the Hamiltonian and momentum constraints to an accuracy of $10^{-14}$.
A computational grid of $323\times 5\times 323$ points is used to cover a computational domain of $(\pi r^2\times z)=(\pi\times 38.5^2\times 77.0)$. 
Using the $\Gamma$-freezing shift and $1+\log$ slicing in a $3+1$ Baumgarte-Shapiro-Shibata-Nakamura (BSSN) formulation (detailed in Ref. \cite{Miller}), 
the evolution of this Gaussian packet system and the behavior of the axisymmetric object formed is then observed. 

The system is in a non-equilibrium state at the initial time, and a single Gaussian packet is found to oscillate about its corresponding equilibrium TOV configuration.
However, we check that this oscillation time scale is much larger than the time scale of the merging of two such Gaussian packets and their collapse, 
such that the effect of the non-equilibrium nature of the initial data can be considered negligible. 
Furthermore, we note that the non-equilibrium state of the Gaussian packet system reinforces the point of universality
we seek to address in this work. Stated differently, even systems in non-equilibrium states are attracted to the same critical solution as that of the NS system. 

\begin{table}[ht]
\centering
\begin{tabular}{ccccc}
\hline\hline
Configuration & $M_B$ & $\rho_c$ & $d$ & $v_z$ \\ [0.5ex]
\hline
NS & 1.64 & 0.000566 & 27.6 & 0.153 \\
1 & 1.49 & 0.000387 & 29.6 & 0.120 \\
2 & 1.52 & 0.000392 & 29.6 & 0.100 \\
3 & 1.60 & 0.000555 & 27.6 & 0.116 \\
4 & 1.59 & 0.000649 & 27.6 & 0.116 \\ [1ex]
\hline
\end{tabular}
\label{table:t1}
\caption{Initial data configurations for NS and Gaussian packet systems that exhibit critical collapse.
$M_B$ is the total baryonic mass, $\rho_c$ is the NS central baryonic density/height of the Gaussian baryonic density distribution at $t=0$, $d$ is 
the center-to-center separation between the NSs/Gaussian packets, and $v_z$ is the boost velocity of the NSs/Gaussian packets along the 
$z$-direction of the grid.} 
\end{table}

\begin{figure*}
\begin{center}
\includegraphics[scale=0.85]{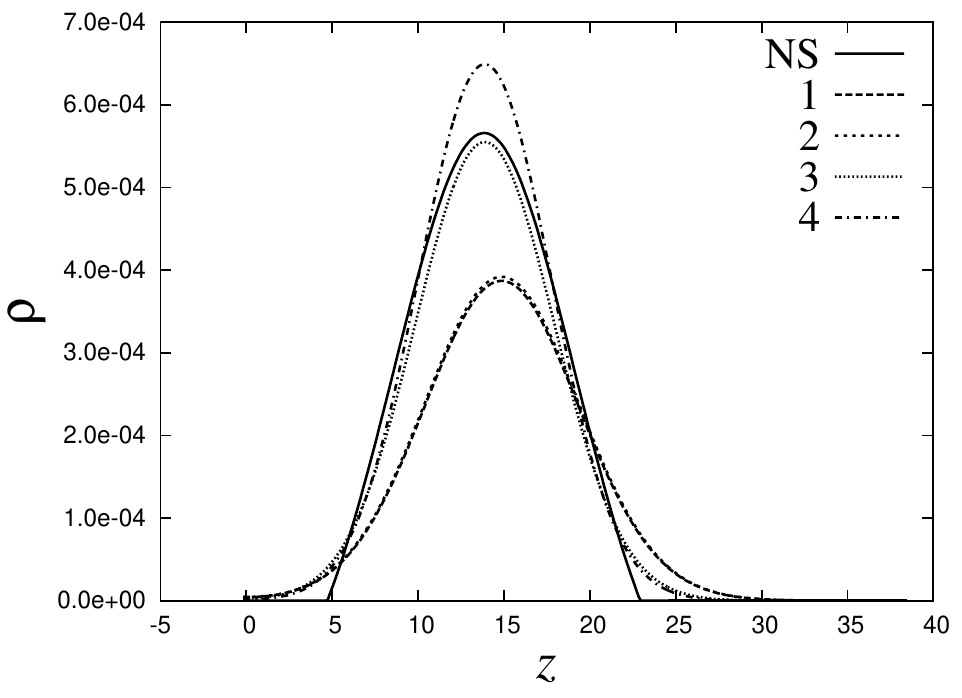}
\caption{Baryonic density distributions for NS and Gaussian packet systems along the direction of head-on collision.
The Gaussian distributions are constructed so as to maintain the baryonic mass of the system as a constant. Therefore, when 
the Gaussian heights are increased, their widths are correspondingly decreased. Configurations 1 and 2 are less compact distributions 
whereas that of 3 and 4 are more compact.} 
\label{fig:fig4}
\end{center}
\end{figure*}   

\begin{figure*}
\begin{center}
\includegraphics[scale=0.8]{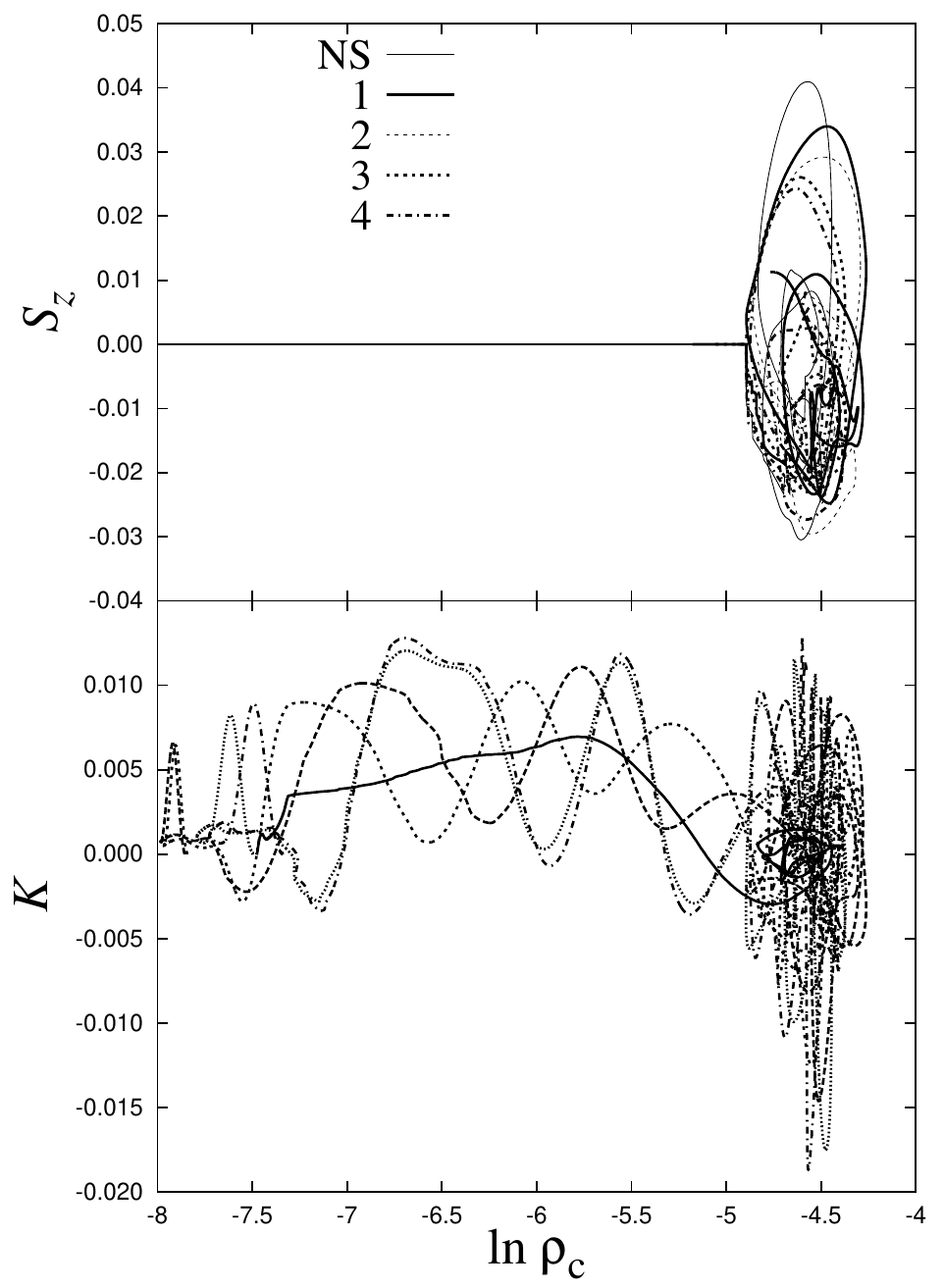}
\caption{Evolution phase diagrams for NS and Gaussian packet systems.
The trajectories of all the configurations start on the left part and are attracted to the basin on the right
part. $K$ denotes the trace of the extrinsic curvature at the coordinate (2.1,0.06,0.06) and $S_z$ denotes the $z$-component of the 3-momentum, $S_i$, of the fluid element
on the $7.5\times 10^{-3}$ baryonic density contour along the $z$-axis, respectively. The natural logarithm is taken of the baryonic density at the center of the grid, i.e.
$\ln\rho_c$, so as to distinguish more clearly the different starting points of the trajectories in the phase diagrams. 
The trajectories in the $S_z$-$\ln\rho_c$ phase diagram start at these different points, but overlap on the $S_z=0$ line up till reaching the basin on the right.}
\label{fig:fig5}
\end{center}
\end{figure*}

Table 1 and Fig.~\ref{fig:fig4} show the configurations that are constructed and their baryonic density distributions along the direction of head-on collision
respectively. Figure~\ref{fig:fig5} shows the phase trajectories for these configurations 
using several evolution variables that form bounded orbits in the attraction basin on the right part of the phase space.
In particular, we note that the correlation between the 3-momentum density of a fluid element away from the center of collision and the
baryonic matter density at the center of collision resembles the Newtonian correlation between the position and momentum of a simple pendulum.
The phase trajectories for the Gaussian packet configurations are compared with that for the NS configuration in Ref. \cite{Jin}. 
We observe that these configurations are all attracted to the same attraction basin in the phase diagrams as that of the NS critical
solution. We propose this as a preliminary suggestion that the NS critical solution is universal 
with respect to various 1-parameter families of initial data, and that the NS critical solution constitutes a semi-attractor. 

\begin{figure*}
\begin{center}
\includegraphics[scale=0.85]{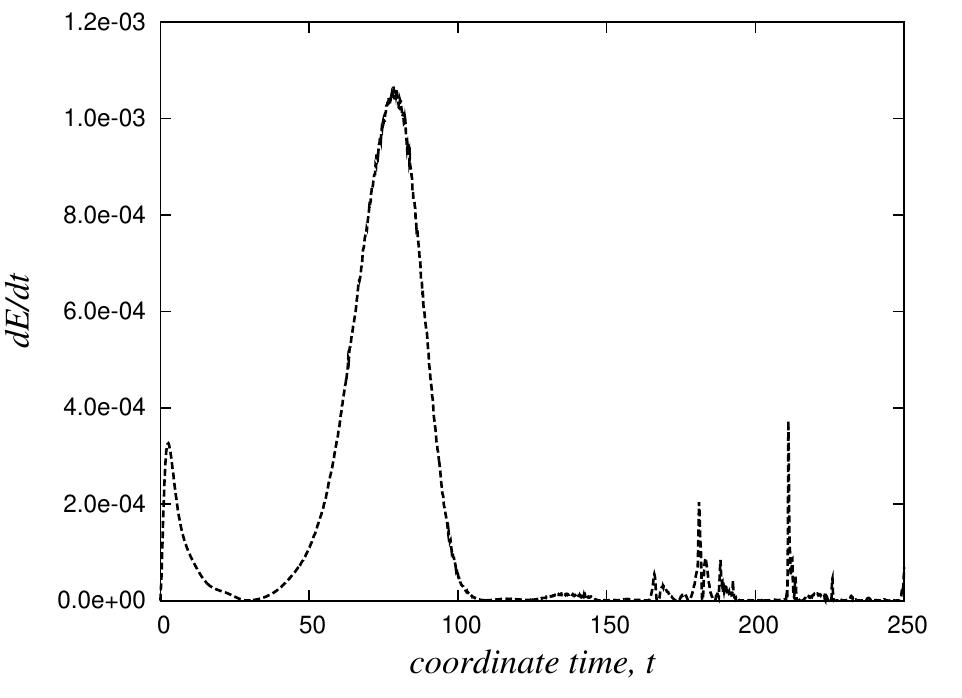}
\caption{Gravitational radiation emission power, $dE/dt$ via the quadrupole formula, Eq. (2), throughout the NS critical solution.}
\label{fig:quad}
\end{center}
\end{figure*}

\begin{figure*}
\begin{center}
\includegraphics[scale=0.6]{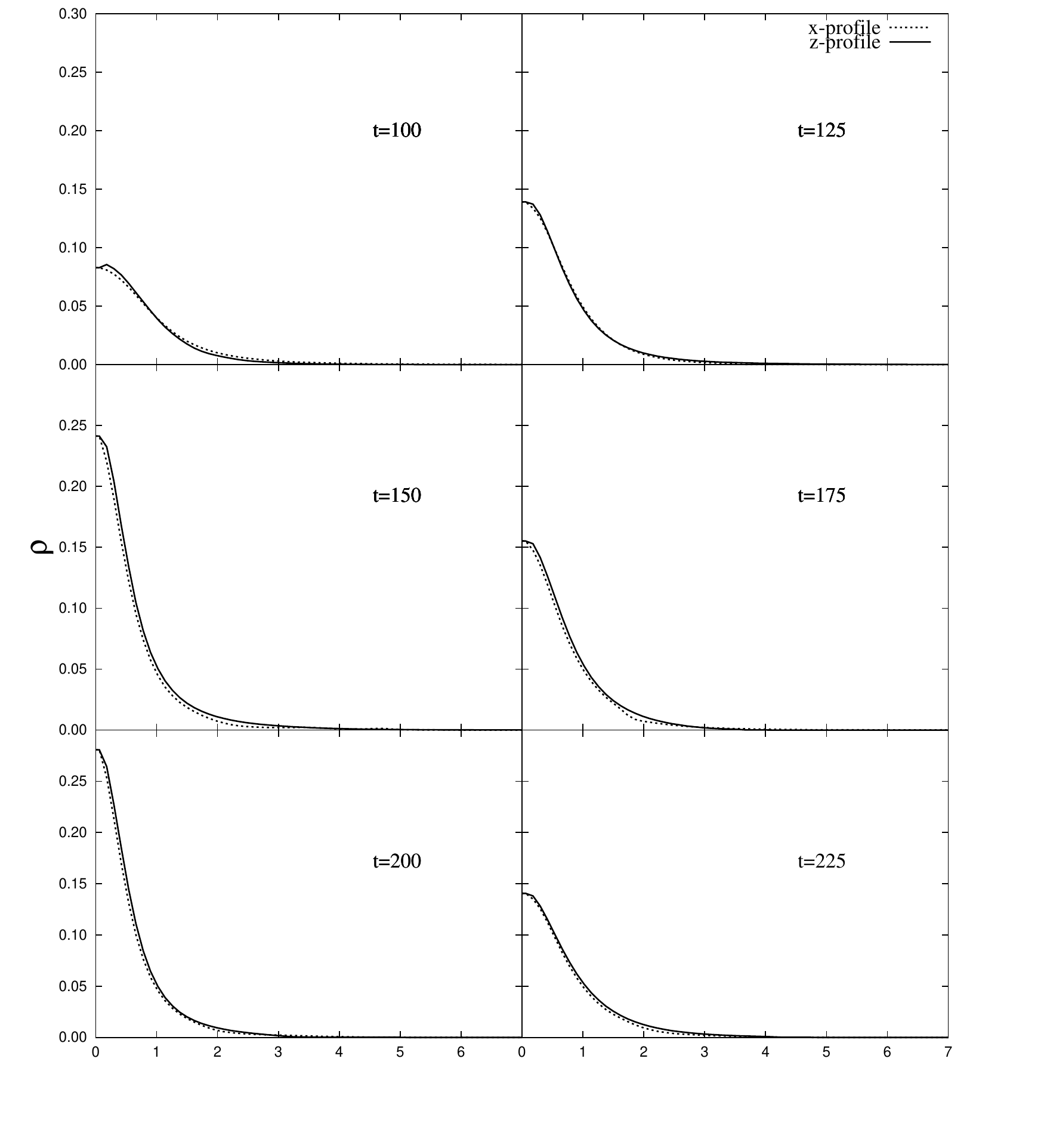}
\caption{Baryonic density, $\rho$ profiles of the merged object along the polar and equatorial directions, i.e., the $x$ and $z$ axes 
respectively, using a sphericity diagnostic spacetime.}
\label{fig:rhoprofSS}
\end{center}
\end{figure*}

\begin{figure*}
\begin{center}
\includegraphics[scale=0.6]{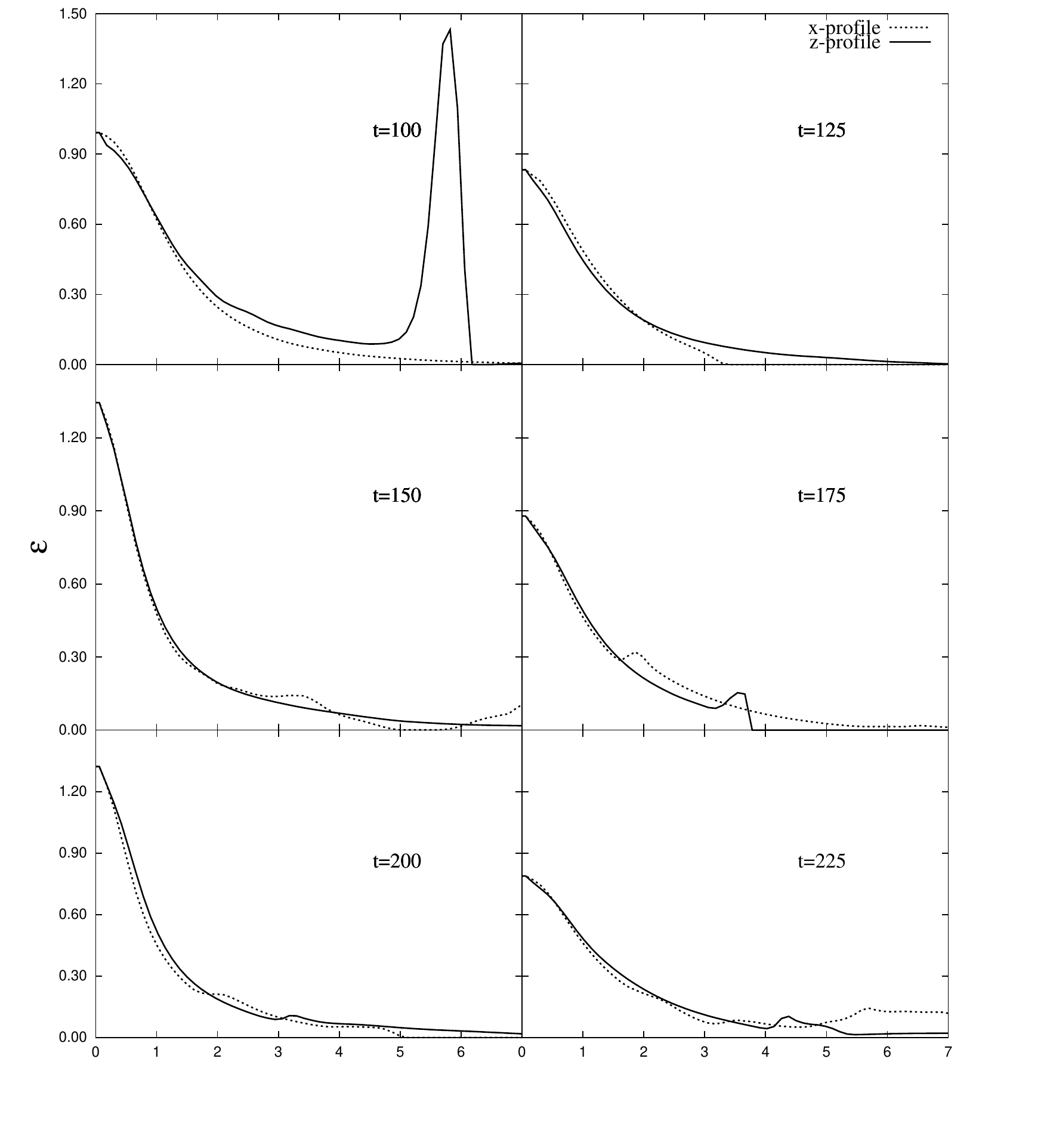}
\caption{Specific internal energy density, $\epsilon$ profiles of the merged object along the $x$ and $z$ axes using the same sphericity diagnostic spacetime.}
\label{fig:epsprofSS}
\end{center}
\end{figure*}

We next measure the power of gravitational radiation emission throughout the dynamical process leading to the semi-attractor state, as well as after the formation of the semi-attractor itself.
Figure~\ref{fig:quad} shows the power that is calculated using the quadrupole formula \cite{Einstein18}:
\begin{equation}
\label{eq:quadeq}
\frac{dE}{dt}=\frac{1}{5}Q^{(3)}_{ij}Q^{(3)}_{ij},
\end{equation}
where $Q_{ij}=\int\rho_e(x^i x^j-\frac{1}{3}\delta^{ij}r^2)d^3 x$ and $\rho_e=\rho(1+\epsilon)$ with $\epsilon$ being the specific internal energy density.
From this figure we see that the power radiated gradually decreases as the system approaches the formation of the semi-attractor at $t\approx 110$,
which is the approximate time of merge of the two NSs.
The power remains at values very close to zero until $t\approx 165$ when it jumps up to positive values that increase with time.
These sharp jumps may be due to noise created by finite-differencing the time derivatives of $Q_{ij}$ \cite{Lin06}.
The average of the power throughout the self-attractor dynamics remains very close to zero.
This suggests that the semi-attractor is undergoing oscillations that are not damped via gravitational radiation,
and thus are likely radial or spherically symmetric pulsations.    

To further affirm the sphericity of the NS semi-attractor in a slicing-independent way, we construct a diagnostic spacetime 
after each successive time interval during the oscillation phase of the critical solution.
The diagnostic spacetime is constructed by setting the lapse function to unity,
the shift vector to zero, and either the $11$- or $33$-component of the $3$-metric to unity depending on whether physical variables along the 
$x$ or $z$ axis are observed \cite{GundlachP}. With these settings, the diagnostic spacetime ensures that 
the hypersurface is perpendicular to the timelike normal unit vector along either the polar or equatorial direction in the axisymmetric spacetime,
and thus being able to eliminate the slicing effect on the oscillations of the object in question along these directions in its spacetime.

Figures~\ref{fig:rhoprofSS} and~\ref{fig:epsprofSS} show the baryonic density and specific heat energy density profiles along the polar and equatorial directions, i.e., $x$ and $z$ directions, on such a diagnostic spacetime at different evolution times during the oscillation phase of the NS critical solution.   
From Fig.~\ref{fig:rhoprofSS}, we observe an almost full overlap of the baryonic density profiles along the polar and equatorial directions
throughout the oscillations of the merged object of the NS critical solution.
Figure~\ref{fig:epsprofSS} shows a similar overlap in the specific heat energy density profiles, with the exception of the first panel 
where there exists a peak in the $z$-profile outside of an otherwise approximate full overlap of both profiles. 
At $t=100$, the specific heat energy density is reflecting a shock wave traveling outward 
in the $z$ direction during the merge of the NSs and thus showing that the system has not fully settled into the semi-attractor state. 
Aside from this, the approximate full overlap of the profiles along the polar and equatorial directions in the succeeding panels
indicate the near-sphericity of the oscillating NS semi-attractor. 
As a spherical configuration whose quasi-periodic oscillations are undamped for a relatively long period of time,  
it is likely a universal configuration in a quasi-stationary state. 

\section{III. Phase Space Properties}

\begin{figure*}
\begin{center}
\includegraphics[scale=0.85]{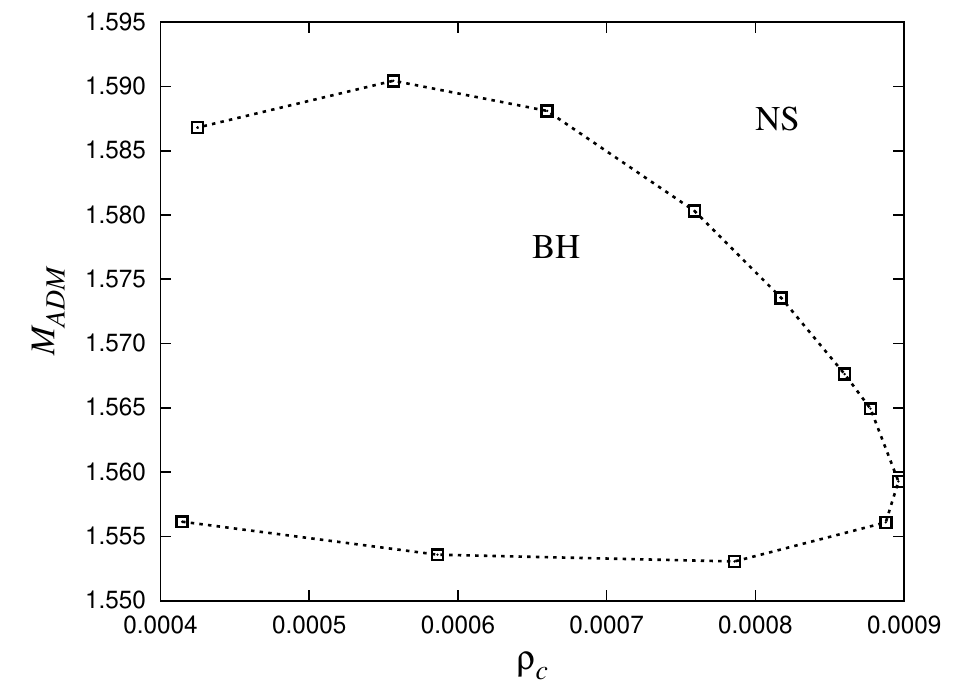}
\caption{The ADM mass-central baryonic density phase space. $M_{ADM}$ denotes the ADM mass and $\rho_c$ denotes the central baryonic density.
BH represents the black hole basin and NS represents the neutron star-like basin. This basin 
representation holds for all the following phase spaces. The dotted line represents the critical surface
that serves as a threshold between the BH and NS basins in this phase space.}
\label{fig:6fT}
\end{center}
\end{figure*}

In this section, we probe the extent of the NS attraction basin using initial data phase spaces.
We note that the newly constructed Gaussian packet initial data possess several additional degrees of freedom as compared to that available for the NS initial data.
These degrees of freedom correspond to the height and width parameters of both the Gaussian density distributions and their $3$-metric functions, 
which can be freely adjusted to yield different initial data when solving the Hamiltonian and momentum constraints.  

We first construct an initial data phase space by maintaining the baryonic mass of the system,
and varying the ADM masses, the heights and correspondingly the widths of the Gaussian density distributions. 
For the system with baryonic mass $M_B=1.6389$ and $d=27.6$, we observe the range of ADM masses and central baryonic densities at which the system exhibits critical collapse (Fig.~\ref{fig:6fT}).
We find that there is a turning point in this phase space at approximately 
$(\rho_c,M_{ADM})=(0.0009,1.56)$, which corresponds to the NS critical solution. An initial data configuration to the left of the 
critical surface collapses to a black hole and one to the right results in an NS-like object. 
This is a somewhat surprising behavior since it implies that with all other parameters fixed, when the compactness of the Gaussian packets is increased 
past the threshold, the system is supported from collapse to a black hole and instead forms an NS-like object. 
Looking into the dynamics, the configuration with higher compactness produces more shock heating than the one with lower compactness. 
This increased shock heating is related to the fact that during the merging process, the matter of the more compact packets spreads out across 
the outer regions of the merger site more than the less compact packets. 

\begin{figure*}
\begin{center}
\includegraphics[scale=0.85]{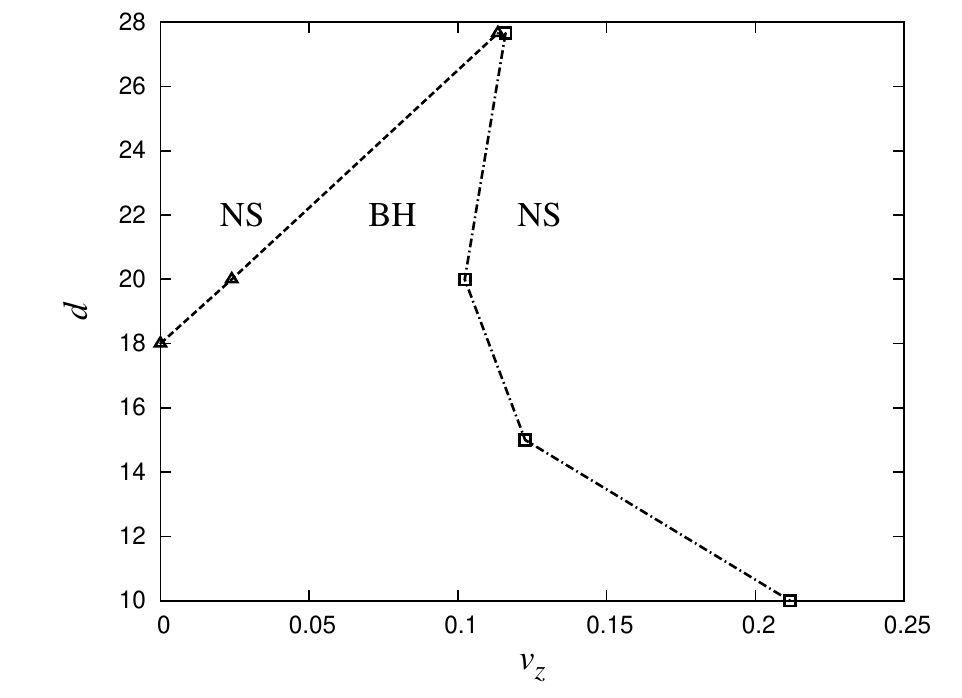}
\caption{The separation distance-boost velocity phase space. $d$ is the center-to-center separation distance between the two Gaussian packets and 
$v_z$ denotes the boost velocity along the direction of head-on collision of the Gaussian packets.
The dashed and dotted-dashed lines represent two different critical surfaces that serve as the NS-BH threshold and the BH-NS threshold respectively
as the boost velocity increases.}
\label{fig:DvV}
\end{center}
\end{figure*}

On the left side of Fig.~\ref{fig:6fT}, the two Gaussian packets are too diffuse to fit within the computational domain used for this phase space. Therefore, the endpoints on the left side of the critical surface here are a direct consequence of the finiteness of the computational domain. 
However, beyond the top, bottom, and right edges of the critical surface, no critical collapse behavior is observed.
Therefore, this critical surface also shows the extent of the NS critical solution attraction basin 
in terms of ADM masses and central baryonic densities at a certain baryonic mass with a polytropic EOS as mentioned in the Introduction.
To further explore the extent of the attraction basin, 
we construct another phase space using the boost velocity and the separation distance between the Gaussian packets.
Figure~\ref{fig:DvV} shows two critical surfaces for the system with $M_B=1.6389$ and $\rho_c=8.935\times 10^{-4}$. 
The critical surface on the left involves the system being at a fixed separation distance passing the threshold from the 
NS-like phase into the black hole phase and passing back into the NS-like
phase through the right critical surface. 
This indicates that when the system is given enough kinetic energy, its collapse is also delayed, similar to what is mentioned in the previous paragraph.

Both the critical surfaces end at $d/2\approx 14$
at the top of the phase space. The bottom extent of the right critical surface is
limited when the separation distance between the two Gaussian packets decreases
to the point where the packets merge to become one single packet. 
The total baryonic mass of this single packet exceeds the maximum baryonic mass of an
equilibrium TOV configuration with the same EOS with $\kappa=80$ and $\Gamma=2$.
Consequently, the system collapses into a black hole even at zero implosion velocity. We
hereafter set the baryonic mass of the system at $1.6$, below the maximum baryonic mass.
In addition, it is known that the maximum point of the baryonic mass-central baryonic density relation for equilibrium TOV star
configurations is \textit{not} a critical point. Therefore, a study on how the transition occurs across the boundary of 
the NS critical solution attraction basin is warranted.
To do this, we increase the central baryonic density of
the Gaussian packets so that it approaches the maximum point, i.e., $\rho_c\approx 4.0\times 10^{-3}$.
For each increase of the central baryonic density, it and all other parameters are then fixed while the implosion velocity is varied.
In this transition, the critical point is seen to move monotonically to lower values of the implosion velocity. 
The oscillations of the critical solution are also seen to decrease in amplitude. The average
of the oscillations in the central lapse function, $\alpha_{avg}$ shifts up whilst that of the central
baryonic density, $\rho_{avg}$ shifts down, approaching resemblance to the normal mode oscillations of a stable equilibrium TOV star configuration. 

At $\rho_c\approx 1.833\times 10^{-3}$, the threshold between the black hole
and NS-like phases approaches the point of disappearing. At central densities beyond this value, the
system oscillates at the normal mode frequencies of its corresponding stable equilibrium
TOV star configuration, for all implosion velocities from $v=0$ up to $v=0.8$, a velocity very near the speed of light.
However, we observe that the transition point across $\rho_c\approx 1.833\times 10^{-3}$ itself is characterized by a critical solution 
between the black hole phase and the NS-like phase. Specifically, with all other parameters fixed, 
when the central baryonic density exceeds this threshold value, the single Gaussian packet settles into an oscillating state
about a corresponding stable equilibrium TOV star configuration. When the central baryonic density is less, it collapses into a black hole.
This behavior is similar to what is observed earlier in Fig.~\ref{fig:6fT}.

\begin{figure*}
\begin{center}
\includegraphics[scale=0.85]{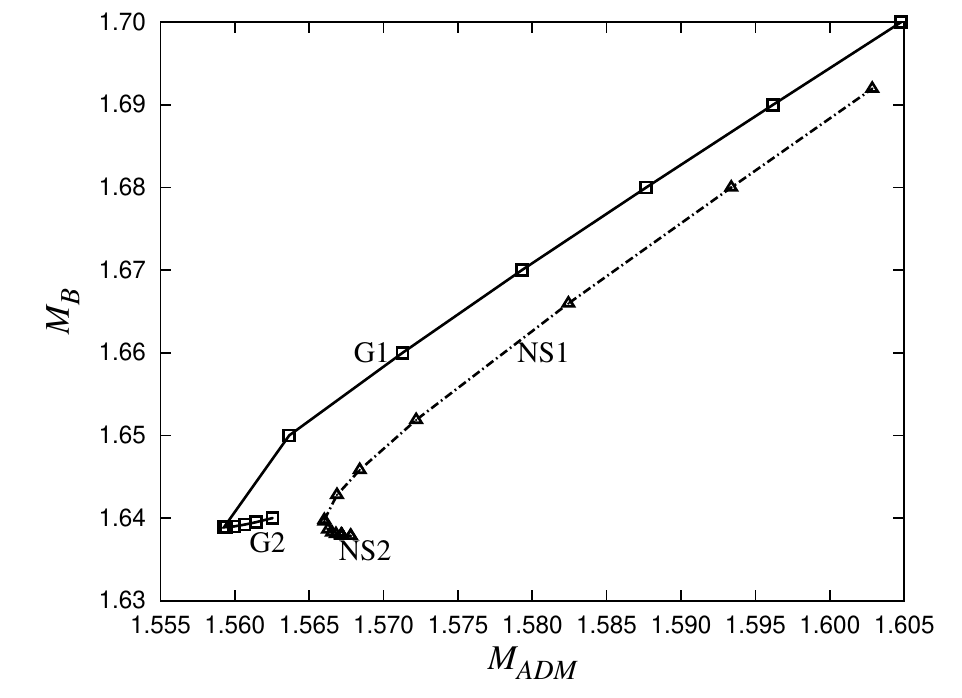}
\caption{The baryonic mass-ADM mass phase space. The solid line represents the critical surface obtained for a Gaussian packet system. 
G1 is the NS-BH threshold whereas G2 is the BH-NS threshold.
The dotted dashed line is the critical surface obtained for the neutron star system. 
Similarly, NS1 is the NS-BH threshold and NS2 is the BH-NS threshold.
Both G2 and NS2 can be extended monotonically further to the right at least up to the baryonic mass value reached by G1 and NS1.}
\label{fig:6gT}
\end{center}
\end{figure*}

We also construct a phase space using the baryonic mass and the ADM mass to compare
with the critical surface found for the NS system (Fig.~\ref{fig:6gT}).
In this phase space, both systems are at set at $d=27.6$.
A turning point behavior, similar to that seen in Fig.~\ref{fig:6fT}, is observed in both the critical surfaces.
We hereby label the ADM mass value where this turning point occurs for the Gaussian critical surface, as $M_t$.
Another important feature is the existence of a common baryonic mass range but a different and overlapping ADM mass range. 
The ADM mass of both systems here possesses a one-to-one correspondence with the boost velocity of the two colliding Gaussian packets or NSs.
Since the choice of gauge or coordinate system affects the value of the boost velocity, we see that the ADM mass changes 
when a different choice of gauge is used. 
The discrepancy in the ADM mass ranges is thus due to the fact that the
Gaussian packet system possesses a coordinate system that differs from the NS system,
where the Gaussian packet system employs a Gaussian function as the $3$-metric of its spacetime and the NS system employs the TOV metric. 

However, we observe that a critical surface with a common ADM mass
range can also be obtained, indicating that the coordinate system of the Gaussian packet system
can be freely adjusted using the additional degrees of freedom mentioned earlier in this section.
In fact, a family of critical surfaces can be obtained for the Gaussian packet system in the baryonic
mass-ADM mass phase space with a common baryonic mass range. These surfaces also possess $M_t$'s that lie within the ADM mass
range indicated in Fig.~\ref{fig:6gT} by the NS critical surface, i.e. between $1.56$ and
$1.6$. As baryonic mass is conserved, the overlap in the baryonic mass range of both critical
surfaces confirms that Gaussian packet/NS-like systems evolve toward the same critical solutions as the NS system.

\section{IV. Critical Index Properties}

\begin{figure*}
\begin{center}
\includegraphics[scale=0.85]{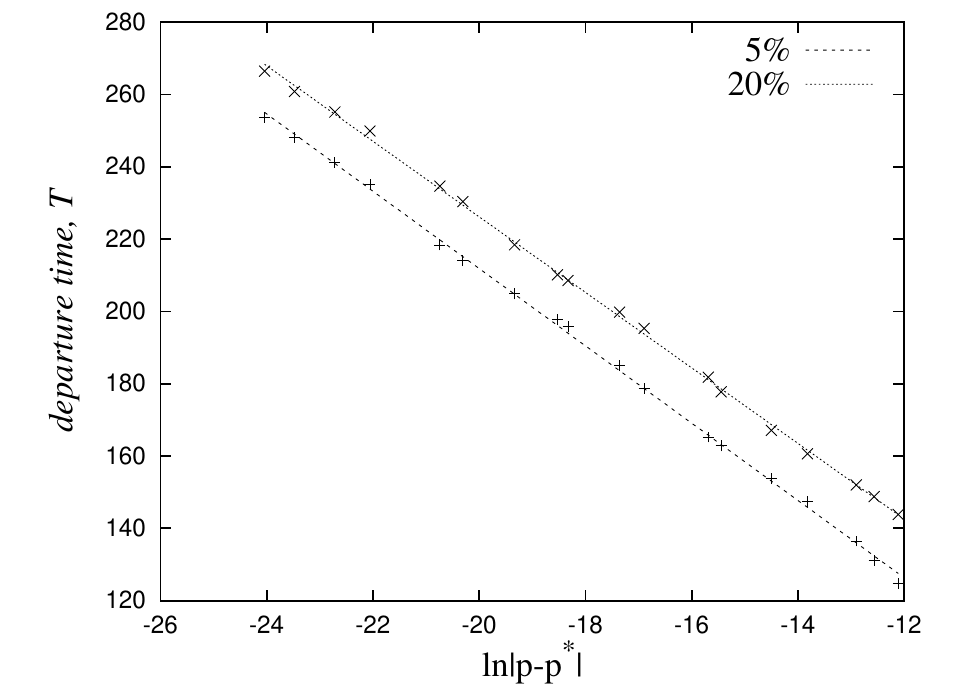}
\caption{Critical index, $\gamma$ extraction for a Gaussian packet system. The critical index is calculated as the slope of a $T$ vs $\ln|p-p^{*}|$ line. $5\%, 20\%$ in the legend label the lines obtained when $T$ is read at departure thresholds, $|(p-p^{*})/p^{*}|\times 100\%=5\%,20\%$ respectively. The fact that the two lines in this figure are parallel implies that the critical index value for this system is within the convergence regime with respect to departure thresholds from $5\%$ up to $20\%$.}
\label{fig:crez5h}
\end{center}
\end{figure*}

\begin{figure*}
\subfigure[]{
\includegraphics[scale=0.7]{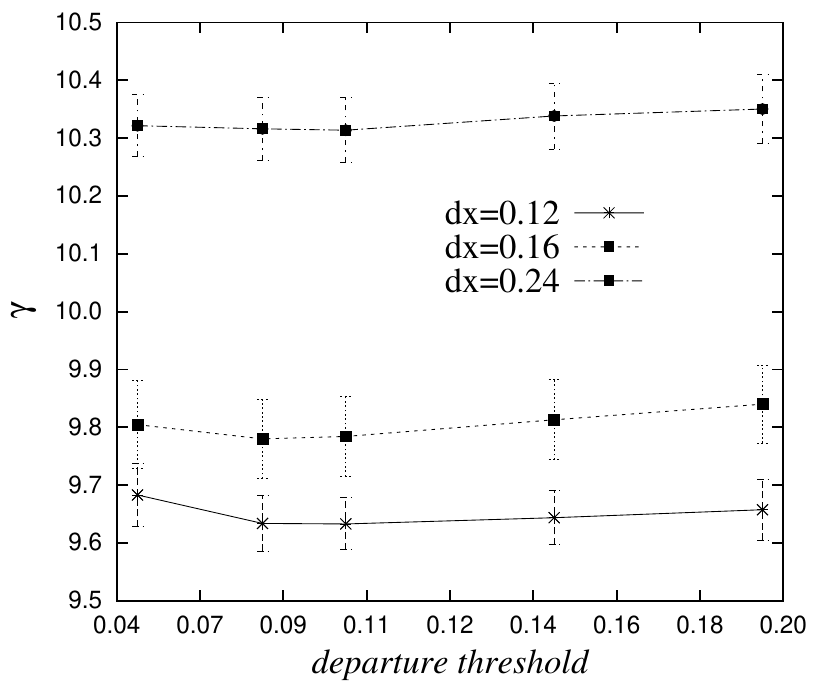}
}
\subfigure[]{
\includegraphics[scale=0.7]{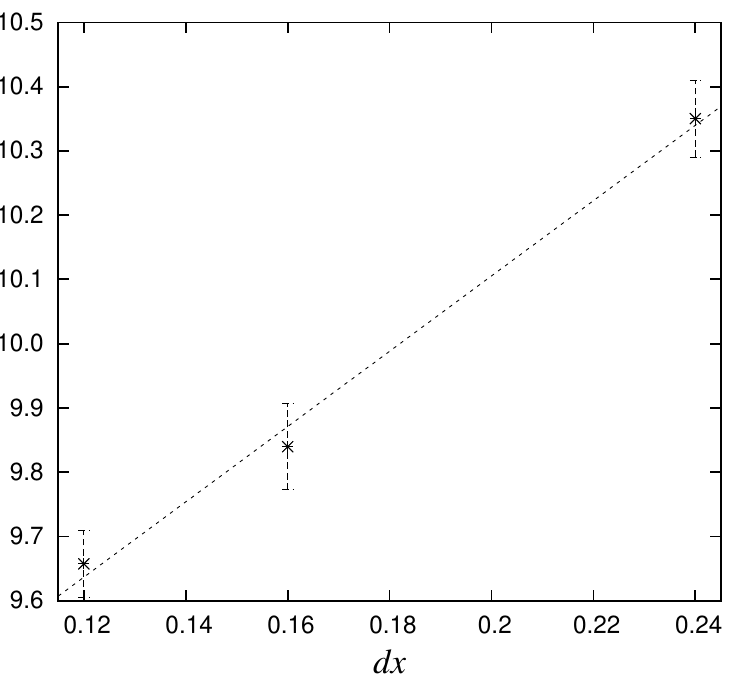} 
}
\caption{(a): Variation of the critical index, $\gamma$ with respect to departure threshold at different grid resolutions, $dx$ for a Gaussian packet system. The critical indices converge to constant values in approximately the same departure threshold range at all three grid resolutions shown. (b): Approximate first-order convergence of the critical index, $\gamma$ with respect to grid resolution for departure threshold $20\%$.}
\label{fig:6kand6m}
\end{figure*}

\begin{figure*}
\subfigure[]{
\includegraphics[scale=0.7]{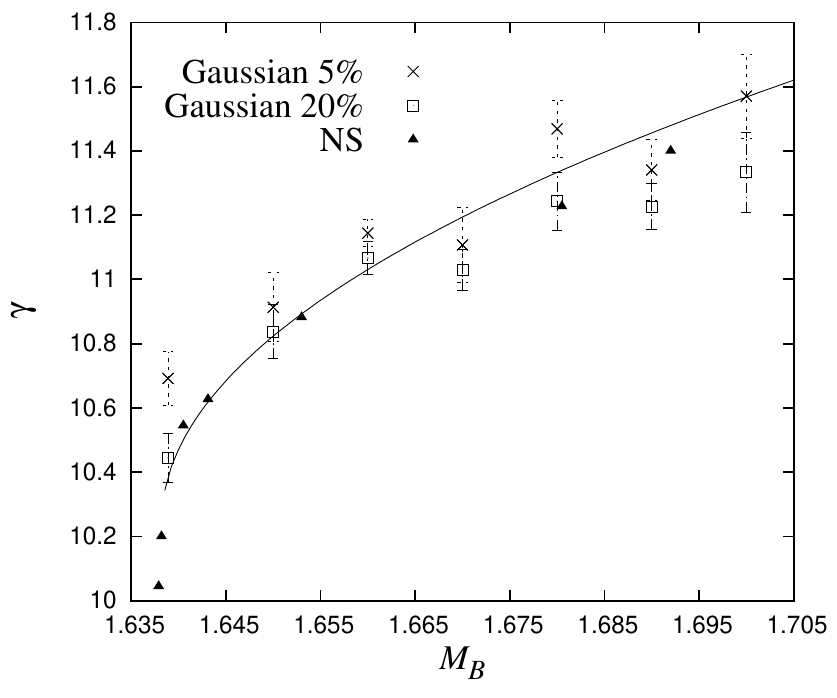}
}
\subfigure[]{
\includegraphics[scale=0.7]{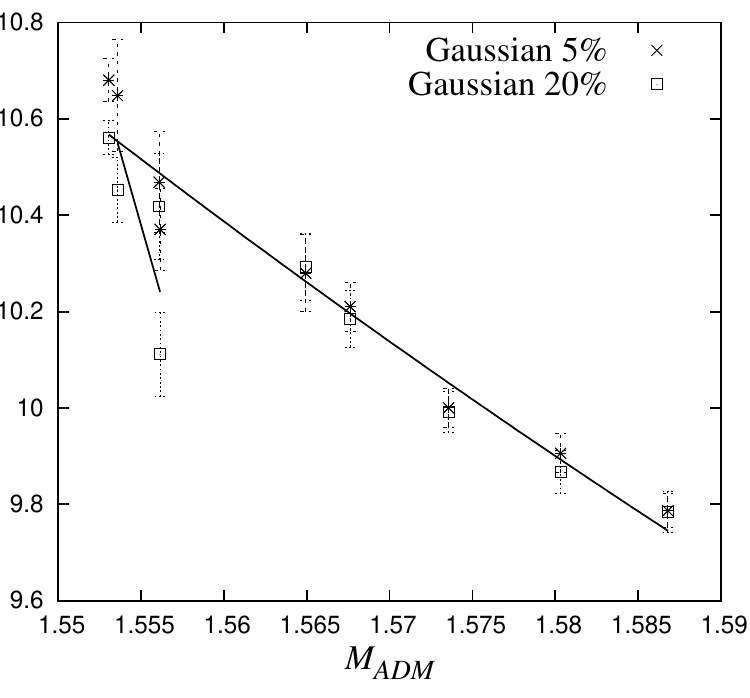}
}
\caption{(a): Comparison of critical indices, $\gamma$'s between the NS and Gaussian packet systems at different baryonic masses. The stars and boxes respectively indicate critical index values calculated for the Gaussian packet system at the departure thresholds mentioned in the legend, as do for all the following figures (Fig.s~\ref{fig:6jTandGAvV}(b) to \ref{fig:convGAandconvGAb}).
The solid line is a fit of the baryonic mass dependence of the critical index for the NS system. (b): Variation of the critical index, $\gamma$ with respect to ADM mass for baryonic mass $1.6389$. The solid line is a fit of the averages of critical indices calculated for the two departure thresholds indicated by the legend.}
\label{fig:6jTandGAvV}
\end{figure*}

\begin{figure*}
\subfigure[]{
\includegraphics[scale=0.7]{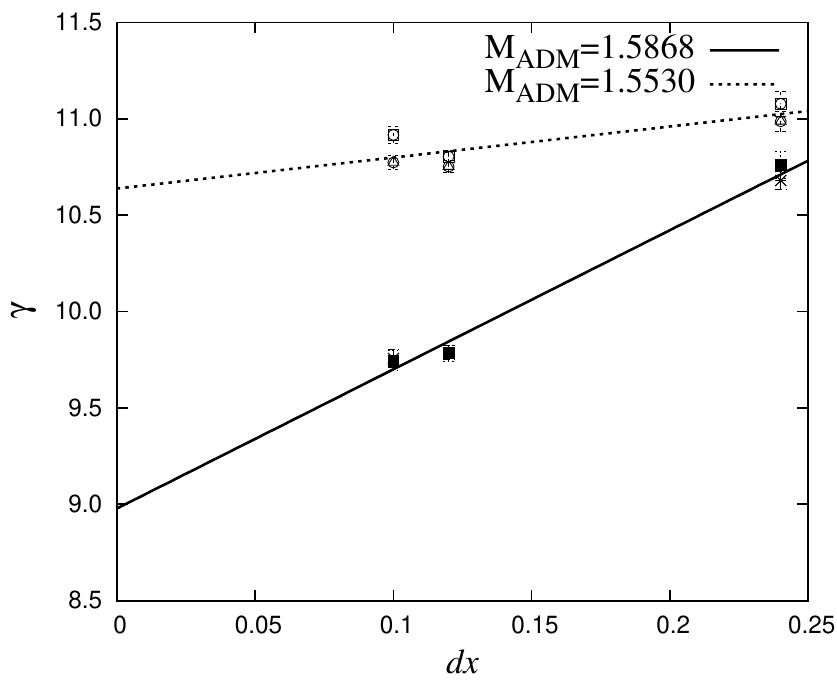}
}
\subfigure[]{
\includegraphics[scale=0.7]{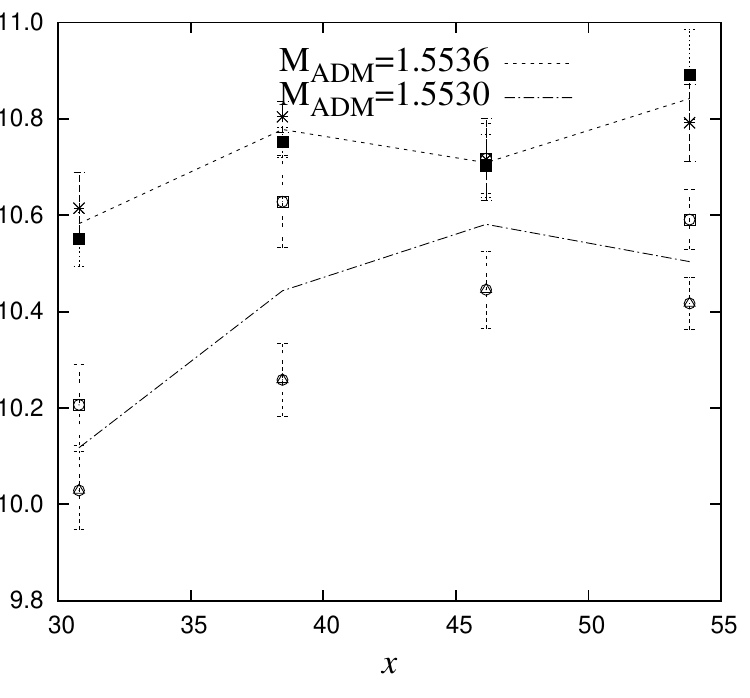}
}
\caption{(a): Convergence of critical indices, $\gamma$'s with respect to grid resolution, $dx$.
The solid and dotted lines are fit lines for the averages of the critical indices calculated at the $5\%,20\%$ departure thresholds, for the respective configurations with ADM masses indicated by the legend. (b): Convergence of critical indices, $\gamma$'s with respect to boundary location, $x$. The dashed and dotted-dashed lines pass through the averages of the critical indices calculated at the $5\%,20\%$ departure thresholds, for the respective configurations with ADM masses indicated by the legend.} 
\label{fig:convGAandconvGAb}
\end{figure*}

Following Ref. \cite{Jin}, the critical index, $\gamma$, is calculated using the construct of the departure time of evolution variables for near-critical solutions. 
When the percent difference between the evolution variable of a near-critical solution 
and that of the critical solution itself reaches $5\%,20\%$ etc, we take the time as departure time, $T_{0.05},T_{0.2}$ etc respectively. 
For the critical index calculations presented here, we take the lapse at the center of collision as the evolution variable.
For each critical point, we find that $T$ is linearly proportionate to $\ln|p-p^{*}|$, where $p$ is any initial data parameter that we choose to vary while fixing the rest, and $p^{*}$ is the initial data parameter for the critical solution. 
In the case of Fig.~\ref{fig:crez5h}, the initial data parameter varied is the boost velocity of the Gaussian packets.
Straight best fit lines are obtained and the standard deviations
of their slopes are plotted as error bars in the following figures (Fig.s~\ref{fig:6kand6m} to \ref{fig:convGAandconvGAb}). 
Other errors that affect the numerical results include the error caused by the finite differencing scheme and
the error of the exponential law of near-critical solutions leaving the critical solution.

Before going into the properties of the critical index,
we check its convergence with respect to the finite-differencing resolution $dx$. Figure~\ref{fig:6kand6m}(a) shows the critical indices for different departure thresholds 
(defined in the Fig.~\ref{fig:crez5h} caption) at three different resolutions for a sample Gaussian system. At the departure threshold of $20\%$, Fig.~\ref{fig:6kand6m}(b) shows an approximate first order convergence with respect to the resolution,
which is in line with the order of convergence expected for the high resolution shock capturing Total-Variation-Diminishing (TVD) hydrodynamics scheme 
being used in our simulations (as detailed in Ref. \cite{Miller}).
The grid boundary used for the simulations with resolution $dx=0.16$, i.e., $41.0$, is slightly larger than that with $dx=0.12$ and $dx=0.24$, which is at $38.5$, due to the restrictions on the number of grid points imposed by the multigrid initial value problem solver. 
It should be noted that due to this, the convergence law may not appear as clean as when the grid boundary is strictly fixed. 

Now we consider Fig.~\ref{fig:6jTandGAvV}(a) which shows the baryonic mass dependence of the Gaussian packet critical index compared with that of the NS.
Here, we observe that the critical indices for both the Gaussian packet and NS systems increase with the baryonic masses of the systems.
This indicates that the time scale of departure from the critical solution decreases as the baryonic mass of the system increases.
The overlap of the critical indices of both the Gaussian and the NS systems shows that the NS semi-attractor is universal at least with respect to 
these two different types of initial data. Figure~\ref{fig:6jTandGAvV}(a) specifically shows that 
for the same baryonic mass and within a small interval of ADM masses, the Gaussian system produces critical indices that lie 
within the error bars of those produced by the NS system.

With this observation, we next attempt to determine whether the critical index has a $1$- or $2$-parameter dependence.
In particular, we determine whether it solely depends on the baryonic mass, or whether it depends on \textit{both} the baryonic mass \textit{and} the gravitational energy present in the system.
Taking note that the critical surface in the ADM mass-central baryonic density phase space (Fig.~\ref{fig:6fT}) is obtained by fixing the baryonic mass of the system,
we study how the critical indices change along this surface.
Figure~\ref{fig:6jTandGAvV}(b) shows the continuous variation of the critical index along this critical surface
with respect to the ADM mass while fixing the baryonic mass at $1.6389$. The solid line is a fit of the averages of the indices obtained at departure thresholds $5\%$ and $20\%$.
In addition, at $M_{ADM}\approx 1.5525$, the critical surface in Fig.~\ref{fig:6fT} exhibits a slight turn upwards.
The "scatter" in the critical index values into a separate branch on the left of Fig.~\ref{fig:6jTandGAvV}(b) reflects this slight turn upwards of the critical surface. 
With this separate branch, the Gaussian packet system that undergoes critical collapse possesses the same ADM mass for two different compactness values.
In other words, we see a direct one-to-one correspondence between the critical index and the compactness of the system.

It is important to note that this variation of the critical index with respect to the ADM mass for a fixed baryonic mass, is also manifested in the 
NS system. Taking the baryonic mass of $1.65$, we extract the critical index for the critical points on both branches $1$ and $2$ of the NS system critical surface in Fig.~\ref{fig:6gT}.
In accordance with Fig.~\ref{fig:6jTandGAvV}(b), we find a discrepancy between these critical indices that is bigger than their error bars. 
On branch $1$ of the critical surface with this baryonic mass, the critical index obtained is $10.9\pm 0.1$ at departure threshold $5\%$ and $10.84\pm 0.08$
at departure threshold $20\%$, whilst on branch $2$,
the critical index is found to be $10.26\pm 0.06$ at departure threshold $5\%$ and $10.24\pm 0.07$ at departure threshold $20\%$.
Described differently, the critical points on branches $1$ and $2$ of the critical surface represent two different phase thresholds respectively.
These observations in both the Gaussian packet and NS systems, of the dependence of the critical index on the gravitational energy while fixing the baryonic mass, 
suggest that the critical index has a $2$-parameter dependence. 

To provide further verification of this $2$-parameter dependence, we consider ruling out the scenario where for the same baryonic mass, the Gaussian packet systems 
possess the same gravitational energy at late time when the semi-attractor state has been reached.
It is important to consider this scenario since it is commonly known that equilibrium configurations, such as TOV stars, with the same baryonic mass and EOS,
possess at most two different gravitational energies, one for the stable configuration and the other for the unstable configuration.  
We check the values of the baryonic mass and the gravitational energy of the Gaussian packet system after the formation of the merged object
for two configurations on the ADM mass-central baryonic density phase space.
For the configurations with $M_{ADM}=1.5530$ and $M_{ADM}=1.5868$ respectively, 
we look at the variation of these values with respect to grid resolution.
The ADM masses of these two configurations do not correspond to that of any pair of stable-unstable equilibrium TOV configurations
for the polytropic EOS with $\kappa=80$ and $\Gamma=2$.

For the purpose of checking late-time values, we take the averages of the gravitational masses, $\langle M_G\rangle$, measured by:
\begin{equation}
\langle M_G\rangle=\left\langle \int\Psi^5[\rho_e+\frac{1}{16\pi}(\tilde{A}_{ij}\tilde{A}^{ij}-\frac{2}{3}K^2-\tilde{R}\Psi^{-4})]\sqrt{\tilde{\gamma}}d^3 x \right\rangle,
\end{equation}
on the finite computational grid over the period of two oscillations in the lapse function at the center of collision.
In this equation, $\Psi=e^{\phi}$ with $\phi$ being the conformal factor in the $3+1$ BSSN split, and $\tilde{A}_{ij}$, $\tilde{A}^{ij}$, $K$, $\tilde{R}$ and $\tilde{\gamma}$
are standard spacetime evolution variables in the $3+1$ BSSN formalism.

At the highest resolution used in this paper, i.e., at $dx=0.1$, 
we observe that the $\langle M_B\rangle$ values for the configurations reach to $1.634\pm 1\times 10^{-4}$ and $1.635\pm 2\times 10^{-4}$ respectively.
since the baryonic masses of the initial data have been set to the same value up to the eleventh decimal, 
we expect the $\langle M_B\rangle$'s to converge to a small value band with a width on the order of $0.001$. 
The baryonic mass is conserved to the level of less than $1\%$ up to the time of formation and first two oscillations of the merged object.
In addition, we note the randomness of the variations beyond the eleventh decimal in the baryonic masses of the different initial data. 
Also, the small variations in the $\langle M_B\rangle$ values as the resolution is increased, do not follow any systematic trend.
On the other hand, the $\langle M_G\rangle$ value for the more compact configuration (with ADM mass $1.5530$) reaches to $1.38\pm 0.001$,
whilst the less compact configuration reaches to $1.41\pm 0.001$. 
With the $\langle M_G\rangle$'s, we clearly see a systematic trend when the resolution is increased, 
where the $\langle M_G\rangle$ for a more compact configuration is consistently and markedly smaller than that for a less compact configuration. 
 
Convergence studies of the critical indices are then performed
to determine the robustness of the $2$-parameter dependence of the critical index in terms of numerical accuracy. 
We test the convergence with respect to grid resolution and boundary location 
for the critical indices of two different configurations on the Fig.~\ref{fig:6fT} phase space.
Both grid resolution and boundary location affect the Hamiltonian and momentum constraint violations during the evolution in our simulations.
Hence, convergence tests are carried out for grid resolution and boundary location separately so as to isolate the numerical errors caused 
by each of these factors respectively.  

For the same two configurations mentioned in the previous paragraph, Fig.~\ref{fig:convGAandconvGAb}(a) shows the convergence of the critical indices with respect to grid resolution 
at a fixed boundary location.
Within the capacity of the computational resources available to us,
the grid resolutions of $dx=0.1$, $dx=0.12$ and $dx=0.24$ are used in this figure to ensure that the boundary location is fixed at $x=38.5$.
They converge to different values, similar to the $\langle M_G\rangle$'s, in roughly a first-order manner,
in line with the order of convergence with respect to grid resolution seen earlier in Fig.~\ref{fig:6kand6m}(b).
Since the fit lines in Fig.~\ref{fig:convGAandconvGAb}(a) exhibit a diverging character when resolution tends to infinity, 
we expect that in the limit of infinite resolution, 
the critical indices for different ADM masses, deviates significantly beyond their error bars. 

Figure~\ref{fig:convGAandconvGAb}(b) shows the convergence of the critical indices for two configurations with ADM mass $1.5530$ and $1.5536$, 
with respect to boundary location at the resolution of $dx=0.12$. 
These are two neighboring configurations on the Fig.~\ref{fig:6fT} phase space, that possess density distributions covered by grid sizes within 
the computational capacity available to us.
The critical indices for even these two neighboring configurations exhibit convergence to values that differ beyond their error bars.
This is indicated as $x$ increases from approximately $38$ onwards, the critical indices averaged from the $5\%$ and $20\%$ departure threshold calculations,
together with their corresponding errorbars, show an approximate asymptotic behavior at different values. 
  
Putting aside the ambiguous physical meaning of $\langle M_G\rangle$, the above observations imply that when a large portion of numerical errors are eliminated,
the critical index depends on both the baryonic mass of the system,
as well as its gravitational energy with its corresponding binding energy present in the physical system.
This indicates that NS semi-attractors can be labeled by both the baryonic mass and the gravitational energy of the system.
Since the oscillating NS semi-attractor is a state that theoretically lasts for an indefinite amount of time, there is a possibility that it is in a type of
time-dependent equilibrium. With multiple ADM masses for each baryonic mass in this probable equilibrium state, 
a type of self-gravitating object previously unknown could exist. However, further work will be needed to rigorously ascertain that the NS semi-attractor is indeed in such time-dependent equilibrium. 
As this paper is nearing completion, an independent study on the nature of the oscillating object formed from a NS-NS collision \cite{Kellermann} is published. There, the authors show that the NS semi-attractor with a certain rest mass and its corresponding ADM mass can be described as an equilibrium TOV star on the unstable branch. The corresponding equilibrium TOV star is taken to have the same polytropic constant, $\kappa$, as the NS semi-attractor.
It is not clear though whether the family of NS semi-attractors found in the current paper can be similarly described.  

\section{V. Conclusions}

In this work, we confirm that the NS critical solution constitutes a semi-attractor
for 1-parameter families of NS-like initial data.
In measuring the power of the gravitational radiation emission throughout the critical collapse dynamics,
and using a sphericity diagnostic spacetime, the object undergoing the oscillations is found to be indeed very nearly spherical,
thus reinforcing the notion that it is a universal configuration.

The phase space properties of the NS critical solution is then analyzed, and interesting behavior is found.
We employ the additional matter and spacetime degrees of freedom available in the new NS-\textit{like} configurations,
to construct phase spaces that otherwise are not possible for the NS system. This is due to the unique characterization of NSs merely 
by their central baryonic densities via the TOV equations.
In these phase spaces, we find turning points with interesting two-threshold features,
indicating that when the kinetic energy exceeds a certain threshold, the collapse into a black hole is delayed and an NS-like object is formed. 
This support from collapse is also seen when the compactness exceeds a threshold.

Apart from these surprising observations, the extent of the critical surfaces in these phase spaces gives an
indication of the size of the attraction basin of the NS critical solution.
For a certain baryonic mass, we determine the range of central densities and ADM masses of configurations that exhibit critical gravitational collapses.
This will help give an approximation of the range of NS or NS-like configurations that can exhibit critical phenomena.
  
Additionally, this work explores the boundaries of the attraction basin
of the NS critical solution by constructing a phase space using the  
separation distance and the ADM mass of the new NS-like system. In decreasing the separation distance, we move
toward an initial configuration that consists of a single NS-like object with a varying implosion velocity.
Fixing the baryonic mass at a slightly lower value than the maximum baryonic mass allowable for a single equilibrium TOV configuration with $\kappa=80$ and $\Gamma=2$,
we find that there exists a boundary, where the threshold approaches the point of disappearing 
when we increase the central baryonic density towards the maximum baryonic mass point.

Finally, via the analysis of the critical index of the NS critical solution, 
we find that the time scale of departure from the critical solution, depends almost proportionately on the baryonic mass of the system. 
With the support of a series of convergence tests, the interesting observation is that
the critical index of the NS critical solution exhibits a dependence on \textit{both}
the baryonic mass and the ADM mass of the system, i.e., a $2$-parameter dependence.
This might be a hint of the existence of a type of time-dependent equilibrium with multiple ADM masses for each baryonic mass.
Further work to prove such time-dependent equilibrium in NS critical gravitational collapses will be required.


\section{VI. Acknowledgements}
I acknowledge Ke-Jian Jin for making accessible the GRAstro-2D code for this work and for providing 
the relevant data for reference. I thank Wai-Mo Suen and Carsten Gundlach for providing important ideas in this work.
I also thank Emanuele Berti and Gentaro Watanabe for critical readings of the manuscript.
The research is supported by the McDonnell Center for Space Sciences, Washington University, and the NSF Grant MCA93S025.
Simulations are performed on the WUSTL Physics Department machines as well as the NCSA Teragrid supercomputing cluster.

\bibliographystyle{prsty} 
\bibliography{references}

\end{document}